\documentstyle[aps,prl,epsfig,multicol]{revtex} 
\begin{document}

\title{Magnetic Scanning Tunneling Microscopy with a Two-Terminal 
Non-Magnetic Tip: Quantitative Results}

\author{T. P. Pareek and Patrick Bruno
\\
Max-Planck-Institute f\"ur Mikrostrukturphysik, Weinberg 2,
D-06120 Halle, Germany}
\maketitle
\begin{abstract}

We report numerical simulation result of a recently proposed
\{P. Bruno, Phys. Rev. Lett {\bf 79}, 4593, (1997)\}
 approach to perform magnetic scanning tunneling microscopy
with a two terminal non-magnetic tip.
It is based upon the spin asymmetry effect of the tunneling current between
a ferromagnetic surface and a two-terminal non-magnetic tip. 
The spin asymmetry effect is due to the spin-orbit scattering in the tip. 
The effect can be viewed as a Mott scattering of tunneling electrons 
within the tip.
To obtain quantitative results we perform numerical simulation
within the single band tight binding model, using recursive Green 
function method and Landauer-B\"uttiker formula for conductance.
A new model has been developed to take into account the spin-orbit 
scattering off the impurities 
within the single-band tight-binding model. 
We show that the spin-asymmetry effect is most prominent 
when the device is in quasi-ballistic regime and the typical
value of spin asymmetry is about 5$\%$.
\end{abstract}

\begin{multicols}{2}
\section*{I. Introduction}

Imaging the magnetic structures of surfaces down to the atomic level is a
major goal of surface magnetism. Magnetic scanning tunneling microscopy(MSTM) 
provides a way to image magnetic domains on surface. In the conventional
approach magnetic sensitivity of tunneling current has been based upon the 
spin-valve effect\cite{Jul}, the tunneling current between two ferro-magnets separated by a tunnel barrier depend on the relative orientation of the magnetizations of the
ferro-magnets. In this approach a magnetic tip has to be used. The
experimental realization of magnetic scanning tunneling microscopy 
based on spin-valve effect was realized by Wiesendanger {\it et. al.}
\cite{Wies},
who investigated a Cr(001)
surface with a ferromagnetic CrO$_2$ tip, 
their observation confirmed the model of topological 
antiferromagnetism between ferromagnetic terraces 
separated by monoatomic steps. They measured a spin 
asymmetry of the order of 20$\%$. 
Recently this method have been used to image magnetic 
domains \cite{wulf,blug,bode,bode1}. 
It was shown that, by periodically changing the 
magnetization of tip, it is possible to separate 
spin-dependent tunnel current from the 
topographic dependent current and hence the 
magnetic structure of surface can be recorded.
Using this method Wulfhekel {\it  et. al.}\cite{wulf} 
studied magnetic domain structure 
on single crystalline Co(0001) surface and 
polycrystalline Ni surface. In Refs. 
\cite{blug,bode,bode1}, 
a two dimensional anti-ferromagnetic structure
of Mn atoms on tungsten(110) surface was investigated.
It was shown that the spin-polarized
tunneling current is sensitive to the magnetic superstructure,
and not to the chemical unit cell \cite{blug}.

However the MSTM with a magnetic tip has a drawback that the magnetostatic interaction between the tip and magnetic sample can not be avoided, which are 
likely to influence the domain structure. In view of this 
an alternative approach was recently proposed to perform the magnetic scanning
tunneling microscopy with a two terminal non-magnetic tip\cite{bruno}. 
It is based upon Mott's
spin-asymmetry effect in scattering caused by disorder\cite{Mott}.
It was shown that due to spin-orbit coupling the tunnel conductance
between the ferromagnetic surface and one of the tip terminal depends on the orientation of magnetization. Because of the spin-orbit interaction
the intensity of scattered beam depends on the orientation of spin-polarization
axis of the incidents electrons, i.e., it is sensitive to the
spin component perpendicular to the scattering plane. In other
words tunnel conductance is spin asymmetric. However to observe
this spin asymmetry effect,caused by Mott scattering, a three terminal
device is a prerequisite. 
Due to the Casimir-Onsager symmetry relation the conductance of
a two-terminal device has to be 
symmetric with respect to 
magnetic field (in our case spin plays the role of magnetic field since as
far as time reversal properties are concerned 'spin' and 'magnetic field are
equivalent), this is a requirement imposed by the underlying microscopic time
reversal symmetry.
However in case of three terminal device,
there is no such restriction on the conductance rather a more
generalized  symmetry relation exists involving all
terminals as shown by B\"uttiker\cite{but}.
Hence to perform  magnetization sensitive
scanning tunneling microscopy with a non-magnetic tip, it is necessary to
use a two terminal tip\cite{bruno}.   

In this work we report numerical simulation result of the three terminal STM
device within the single-band tight-binding model, using recursive Green 
function method and Landauer-B\"uttiker formula for conductance
\cite{datta}.
We have developed a new model to take into account the spin-scattering 
within the single-band tight-binding model.

The paper is organized as follows, in the next section we introduce the
single-band tight-binding model including the spin-orbit interaction
and the three terminal STM device. Section 3 briefly
describes the method of calculation and in section 4 we presents some 
numerical results and discussion.

\section*{2. Model and Method}

A cross section of the system in $xy$ plane,
for the calculation of spin sensitivity of the proposed
two terminal non-magnetic tip is shown in Fig.1. The system consist of
three regions, ({\em i}) the ferromagnetic lead (labelled 1 in Fig.1) , 
({\em ii}) the central region and,
({\em iii}) the  two non-magnetic terminals (labelled as 2 and 3 in Fig.1).
The central region 
is composed of
an insulating tip, such as those routinely used to perform 
atomic force microscopy, coated on two opposite faces 
by a thin metallic film. The metallic
coating has thickness {\it d}.
This is shown in central region where
the empty circles depict vacuum,
black circles corresponds to insulating sites and the rest corresponds to
metallic sites(hatched circles)
and the 
impurities (stars).
Between the ferromagnetic surface (gray circles in Fig.1)
and the tip there is a 
vacuum layer of one lattice spacing (empty circles in Fig.1).  
The tip is placed symmetrical with respect 
to {\it xz} plane.
Current flows along the two faces of tip which makes an angle of
$\pm45$ degrees with the { \it x}-axis. 
The structure shown in Fig.1 consists of three
semi-infinite leads ($-\infty\leq i\leq 1$ and 
$N_x+1\leq i\leq \infty$) separated by tip region $ 1 \leq i \leq N_x $.
The thickness of metallic coating on the tip is $da$ where $a$  being the
lattice constant and the cross section of 
the system is $(N_{y}a\times N_{z}a)$, 
where
$N_y$ and $N_z$ are number of sites along y and z axis. 
For numerical calculation we have taken $N_y$=$N_z$=20 , $N_x$=10
and the metallic coating on the tip has
a thickness of 4 lattice spacing , i.e.,{\it d=4} as shown in Fig.1.

We model the system shown in Fig.1 as a single-band tight-binding Hamiltonian
with nearest neighbor hopping parameter $t$.
To obtain appropriate form of single-band tight-binding Hamiltonian 
including spin-orbit interaction, we discretize the following
single-band Hamiltonian in continuum, on a simple-cubic lattice,  
\begin{equation}
H =\frac{\hat{P}^{2}}{2m^{*}} + V({\bf r}) +
\frac{\Delta}{2}\vec{\mu}({\bf r}) \cdot\vec{\sigma} +
\lambda(\nabla {V}({\bf r})\times\vec{\sigma})\cdot \hat{P} ,
\end{equation} 
\noindent where the first two terms are usual kinetic and potential
energies while  the third and forth terms represent exchange and
spin-orbit interaction, respectively, $m^{*}$ is the effective
mass of electron, $\Delta$ the exchange splitting, $\vec{\mu}$ a
unit vector in the direction of magnetization of FMs and is given by
(cos$\phi$sin$\theta$, sin$\phi$sin$\theta$, cos$\theta$),
$\vec{\sigma}$ the Pauli operator and $\hat{P}$ the momentum
operator.
The discretize form of Hamiltonian reads,
\begin{eqnarray}
H &=&\sum\limits_{\bf r,\sigma,\sigma'}(\varepsilon_{\bf
r}\delta_{\sigma\sigma'}+ \displaystyle\frac{\Delta_{\bf
r}}{2}\vec{\mu}_{\bf r}\cdot\vec{\sigma}_{\sigma\sigma'})c_{\bf
r,\sigma}^{+} c_{\bf r,\sigma'} \nonumber\\
&+& t\sum\limits_{<\bf r, \bf
r'>\sigma} c_{\bf r,\sigma}^{+}c_{\bf r',\sigma}+ H_{so} ,
\end{eqnarray}
\noindent where $H_{so}$ is expressed as

\begin{eqnarray}
H_{so} = -i\alpha_{so} \sum_{\bf r,
\sigma,\sigma'\atop i,j,k,\nu\gamma}\hspace{-0.2cm}
\nu\gamma \Delta\varepsilon_{\bf r+\gamma a_{k},
\bf r+\nu a_{j}}c_{\bf r, \sigma}^{\dagger}c_{\bf r+ \nu a_{j}+
\gamma a_{k}}\nonumber \\
\sigma_{\sigma\sigma'}^{i} \epsilon_{\bf ijk} .
\end{eqnarray}

\noindent Here $c_{\bf r, \sigma}^+$ the creation operator of an electron
with spin $\bf{\sigma}$ at site $\bf r$, $\varepsilon_{\bf r}$
on-site energy and $\Delta\varepsilon_{\bf r+ \gamma a_{k},\bf r+ \nu a_{j}}
= \varepsilon_{\bf r+ \gamma a_{k}}- \varepsilon_{\bf r+ \nu a_{j}}$, 
${\bf a_{i}}$ is the lattice basis vector along axis {\it i},
,
$\sigma_{\sigma\sigma'}^i$, denotes
the Pauli
matrix elements, $\alpha_{so}$ is dimensionless spin-orbit parameter. 
The dummy indice $\nu \gamma$ takes the values $\pm$.The
summation $<{\bf r},{\bf r'}>$ runs over nearest neighbor sites. 
The symbol $\epsilon_{\bf ijk}$ is the Levi-civita's totally
antisymmetric tensor,
where {\it ijk} label the three coordinate axis.

The tight-binding parameters in equation (2) and (3) are related with the
parameters in equation (1) in the following way,
\begin{equation}
t = -\frac{\hbar^{2}}{2m^{*}a^{2}} ,
\end{equation}
\begin{equation}
\alpha_{so} = \frac{\lambda\hbar}{a^{2}} .
\end{equation}

The above tight binding model includes two factors:
spin-dependent band structure and spin-independent disorder. The band structure
takes into account of the difference in the density of states and the Fermi
velocity between the two spin component in the ferromagnet. The disorder represents the structural defects in the real STM tip and is source of
spin-orbit scattering and it takes the form of spin independent 
random variation
in the atomic on-site energies.
In presence of disorder, spin-orbit coupling term  causes hopping along the
diagonal and is the source of spin-flip scattering. 
In this sense this model is equivalent to next-nearest-neighbor (nnb)
tight-binding model, except that in the usual nnb tight-binding model,
hopping amplitude to the next nearest neighbor is fixed while in our model 
it depends on disorder strength and the spin of electron. Hence within this
model spin-relaxation length is determined by disorder strength.

\section*{3. Theory}
As shown in Fig.1, the ferromagnet, the left face of the tip and the
right face of the tip is connected to three reservoirs at chemical potentials
$V_1$, $V_2$ and $V_3$ respectively. Let $I_1$, $I_2$ and $I_3$ be
the corresponding incoming currents in the three terminals
\cite{but,datta}.
\begin{figure}
\begin{center}
\mbox{\epsfig{file=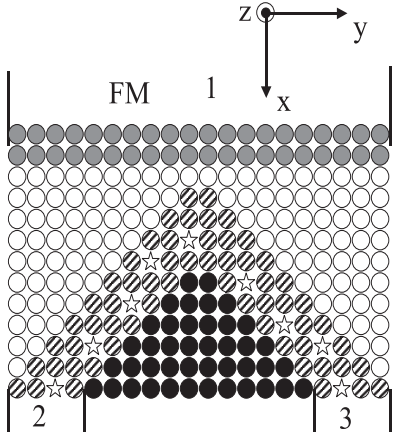,width=3in,height=2.5in,angle=0}}
\end{center}
\caption{ Cross section of tip-geometry shown in $xy$ plane.
Gray circles denote Ferromagnetic sample, the empty circles depict vacuum,
black circles corresponds to insulating sites and the rest corresponds to
metallic sites(circles with downward diagonal lines)
and the 
impurities (empty stars) in the tip.}
\label{Fig. 1}
\end{figure}

The currents are related to potentials by
\begin{equation}
I_{p}=\sum_{q\neq p}G_{pq}(V_{p} - V_{q}) .
\end{equation}
The above expression is gauge invariant and the currents conservation law
$\sum_{i}I_{i}=0$ requires that $G_{pq}$=$G_{qp}$ be satisfied.

The calculation of the conductance of the structure is based upon the 
non-equilibrium Green's function formalism\cite{Kad,Kel}.
When applied to multi-terminal 
ballistic mesoscopic conductor we obtain the 
following result for the conductance \cite{datta}

\begin{equation}
G_{pq}=\frac{e^2}{h} Tr[\Gamma_q G^{R} \Gamma_p G^{A}] .
\end{equation}

Here $p$ and $q$ enumerates the three terminals and the upper indices
R and A refer to
the retarded and advanced Green function of whole structure
taking leads into account. Here $\Gamma_{p(q)} $
self-energy function for the isolated ideal leads and are given by
$\Gamma_{p(q)}$=$t^{2}A_{p(q)}$, where $A_{p(q)}$ is the 
spectral density in the respective lead when it 
is decoupled from the structure. The trace is
over space and spin degrees of freedom, and all the
matrices in equation (4) are of size $(2N_y \times N_z,2N_y\times N_z)$, 
where $N_y$ and
$N_z$ are number of sites along {\it y} and {\it z} 
direction and the factor 2 
takes into account the spin degree of freedom. All the quantities
in the above equations are evaluated at the Fermi energy. To calculate 
the required Green function we use the well known recursive
Green function method \cite{barg}.

\section*{4. Results and discussion}

In this section we present numerical results for a system of cross
section (20 $\times$ 20) in {\it yz} plane and a 
length of 10 lattice spacing along the
{\it x} direction. Number  of metallic  layers on the tip, i.e., {\it d}
in Fig.1 is taken to be 4 lattice spacings. 
The hopping parameter,t ,is same for all pairs
and set to $-1$ for numerical calculation. The on-site energies in the leads
and on the metal coating on the tip is set to be zero, while in the vacuum
layer it is $\epsilon_{vac}=4.0|t|$, and in the insulating region in the tip
it is $\epsilon_{ins}=10.0|t|$. The Fermi level throughout the calculation 
is kept fixed at
$\epsilon_f=3|t|$ above the bottom of the band. For disorder we consider
Anderson Model in which a random on-site energy, characterized by square 
distribution of width {\it W}, is added to the on-site energy of perfect case.
In our case disorder is added only in the metallic coating on the tip;
everywhere
else the system is perfect.

Before we go over to the discussion of  our results 
we briefly mention the  correspondence between 
the physical parameters and the model parameters.
The relevant physical parameters are  mean-free-path, spin-relaxation length 
Fermi Energy and the spin polarization of the ferromagnet at the Fermi level,
and the model parameters, are on-site energy, hopping energy, 
exchange splitting and spin-orbit coupling parameter. 
Physical parameters are related to the model parameters
in the following way,

\begin{equation}
l_{m}=\frac{|t|}{\pi}\sqrt{\frac{\epsilon_f}{|t|}}\frac{1}{N_{3D}(\epsilon_f)
\langle(V- \bar{V})^2\rangle_c} a ,
\end{equation}

\begin{equation}
l_{so}=l_{m}\sqrt{\frac{\tau_{so}}{\tau_m}}
\equiv \frac{3 l_{m}|t|}{2|\alpha_{so}| {\epsilon_f}} ,
\end{equation}

\begin{equation}
P=\frac{N^{\uparrow}(\epsilon_f)-N^{\downarrow}(\epsilon_f)}
{N^{\uparrow}(\epsilon_f)+N^{\downarrow}(\epsilon_f)}
\equiv \frac{\sqrt{\epsilon_f + \Delta} - \sqrt{\epsilon_f - \Delta}}{\sqrt{\epsilon_f + \Delta} + \sqrt{\epsilon_f - \Delta}} ,
\end{equation}

\noindent where $l_m$, $l_{so}$ and P are elastic mean free path, spin relaxation length
and spin polarization of the ferromagnet respectively.
Here $a$ is lattice spacing and $\langle ... \rangle_c$ represents
the configuration averaging, other symbols have the same meaning as defined in
the section 2. Below we present some numerical results 
for one typical realization of the disorder , we have not performed disorder
averaging.

In Fig.2 we  have plotted the  conductance $G_{12}$ and $G_{13}$ as
a function of magnetization angle $\theta$ with respect to {\it z} axis.
We rotate the magnetization in {\it yz} plane such that the
magnetization is always perpendicular to the x-axis
or in other words the angle $\phi$ does not change and has fixed value of 90
degrees.
To be specific, when $\theta$=0 and $\phi=90$ degrees, 
magnetization is parallel to $z$-axis while for $\theta$=90  and $\phi$=90 
degrees
the magnetization is parallel to y-axis.
We have taken $\epsilon_{f}=3|t|$, 
$\alpha_{so}=0.02$ and $\Delta = 2.4|t|$ and the
Anderson disorder strength is $W=1|t|$.
\vspace*{0.8in}
\begin{figure}
\begin{center}
\mbox{\epsfig{file=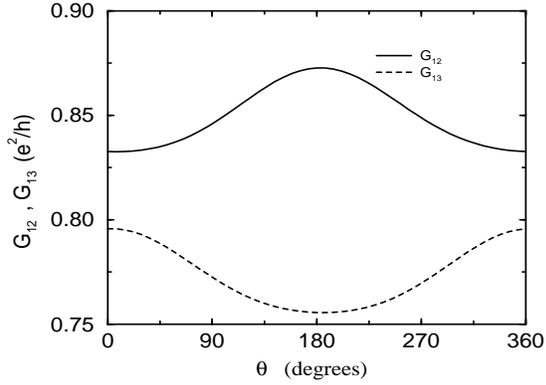,width=2.8in,height=1.5in,angle=0}}
\end{center}
\caption{Fig.2. Conductance ($G_{12}$ and $G_{13}$) versus $\theta$ plot for the two terminals for a
fixed value of $\phi$=90.
The other parameters chosen for this figure are $\epsilon_f=3.0|t|$, 
$\Delta=2.4|t|$, $\alpha_{so}$=0.02, and $W=1|t|$. } 
\label{Fig. 2}
\end{figure}

This set of parameters corresponds to  a mean free path  of $l_{m}=80 a$, 
spin relaxation length of $l_{so}=25 l_{m}$ and 
polarization is $P=50\%$. 
We notice
that the conductance shows approximately cos($\theta$) as a function of
angle, which is expected since  in our geometry the tip is placed symmetrically
to $xz$ plane. However because of disorder the effective axis in the
system does not coincide with the chosen spin quantization axis, 
i.e., $z$ axis, and also since the structure
considered is three dimensional so the scattering plane is not fixed
hence the conductance variation with magnetization
angle does not show an exact cosine behavior. Also we notice that the
variation of $G_{13}$ is opposite to that $G_{12}$. This is in agreement with
the underlying microscopic time-reversible symmetry which requires that
the two terminal conductance should be symmetric under time reversal.
\vspace{0.6in}
\begin{figure}
\begin{center}
\mbox{\epsfig{file=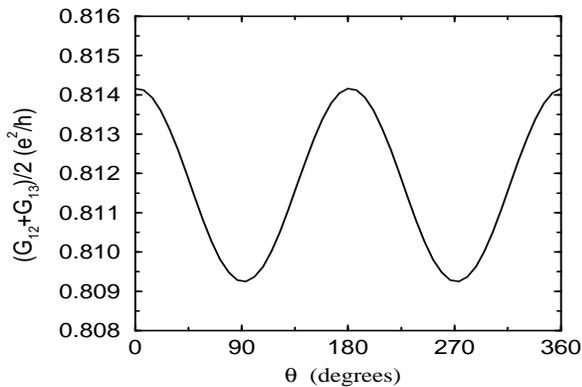,width=3in,height=1.5in,angle=0}}
\end{center}
\caption{Plot of the average two-terminal 
conductance $(G_{12}+G_{13})/2$ versus $\theta$ . The 
other parameters are same as those in Fig.2.  } 
\label{Fig. 3}
\end{figure}

To verify this point in Fig.3 we have plotted, the sum of $(G_{12}+G_{13})/2$,
we see that the two terminal conductance is symmetric with respect to
magnetization angle $\theta$ also the magnitude of oscillation is 
much smaller than either $G_{12}$ or $G_{13}$.
This is due to the fact that the variation of $G_{12}$ or $G_{13}$ with
$\theta$ is of first-order with respect to the spin-orbit coupling, 
whereas the variation of $G_{12}+G_{13}$ is of second order with 
respect to spin-orbit coupling. Actually the latter can be viewed 
as related to the anisotropic magnetoresistance of ferromagnets,
whereas the former is related to the extraordinary Hall effect.
 
In Fig.4 we plot spin asymmetry as a function of polarization of ferromagnet
for terminal 2. 
We have defined the spin-asymmetry as,
\begin{equation}
A=\frac{G^{max}_{12}-G^{min}_{12}}{G^{max}_{12}+G^{min}_{12}}
\end{equation}
where to find $G^{max}_{12}$ and $G^{min}_{12}$ we generate a curve as shown
in Fig.3 for each set of parameters and from those points we get the
corresponding maximum and minimum values. This is necessary
since the variation of conductance with magnetization angle does not follow
exact cosine behavior, the maxima and minima need not to occur 
exactly at zero and 
$\pi$ respectively. 
we have fixed
Fermi energy at $\epsilon_{f}=3|t|$ and $\alpha_{so}=0.02$.
Different curves in Fig.4
corresponds to disorder strengths $W=1|t|$(solid line), $W=2|t|$(dotted line),
$W=2.5|t|$ ( dot-dashed line) and $W=4|t|$ (dashed line) 
corresponding mean-free-paths
are respectively $80 a$, $10 a$, $6 a$, and $3 a$. 
Although all these curves 
corresponds to different
mean-free-paths, however the ratio $l_{so}/{l_m}$ is
same for all the curves and is equal to 25, since this ratio is determined by
Fermi energy and spin-orbit coupling strength, which are kept fixed here.
\vspace{0.6in}
\begin{figure}
\begin{center}
\mbox{\epsfig{file=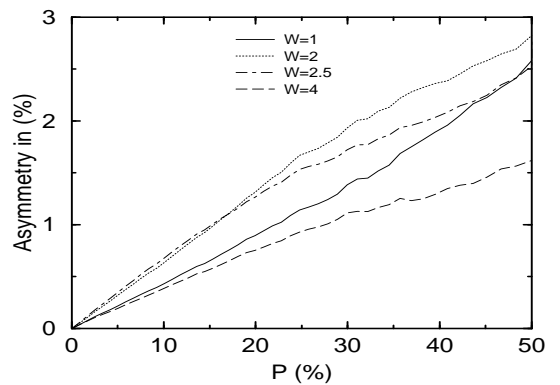,width=2.8in,height=1.5in,angle=0}}
\end{center}
\caption{ Spin asymmetry $A$, as a function of polarization 
for different
disorder strength. The other parameters are
$\epsilon_f=3.0|t|$ and $\alpha_{so}$=0.02. } 
\label{Fig. 4}
\end{figure}

We see that for a fixed disorder 
strength the spin asymmetry 
increases linearly with the polarization or in other words spin asymmetry is
directly proportional to the polarization of ferromagnet.
However for a fixed polarization value, spin asymmetry shows
a non-monotonic behavior. As we increase disorder strength,
spin asymmetry first increase and then starts decreasing. 
This shows that
the spin asymmetry is maximum when the system is in quasi ballistic regime,
since the multiple scattering destroys the spin asymmetry effect.
Which is
clearly visible in the Fig.4 where spin-asymmetry is maximum , for
a fixed value of polarization, at a
disorder strength $W=2|t|$ ,corresponding to a mean-free-path of $10 a$, 
while it is 
minimum for $W=4|t|$ corresponding to a mean-free-path of 
$3 a$ lattice spacings.
The order of magnitude of spin-asymmetry is $5\%$, which is in good
agreement with the  prediction  in  Ref.\cite{bruno}. 

In Fig.5 we have studied the behavior of spin asymmetry as
a function of spin-orbit coupling parameter $\alpha_{so}$.
The other parameters are same as in Fig.4. We notice that
the spin-asymmetry shows a linear behavior for small values of
$\alpha_{so} \leq 0.03$, for larger $\alpha_{so}$ the linear
behavior is no longer seen, because for a fixed disorder, i.e.,
fixed $l_m$, as we increase $\alpha_{so}$, correspondingly ,
$l_{so}$ , i.e. spin-relaxation path decreases, hence the higher
order effect in spin-orbit coupling starts dominating so we
no longer observe a linear behavior. Also we see that 
for a fixed $\alpha_{so}$, spin-asymmetry shows a maximum at
a disorder strength of around $W=2|t|$. This is in harmony with
the results presented in Fig.4. The typical value
of spin asymmetry is of the order of $5\%$. So from the results of Fig.4
and Fig.5 we can say with confidence that the efficency of
the proposed three terminal STM device would be maximum when the
device operates in quasi-ballistic regime.
\vspace{0.6in}
\begin{figure}
\begin{center}
\mbox{\epsfig{file=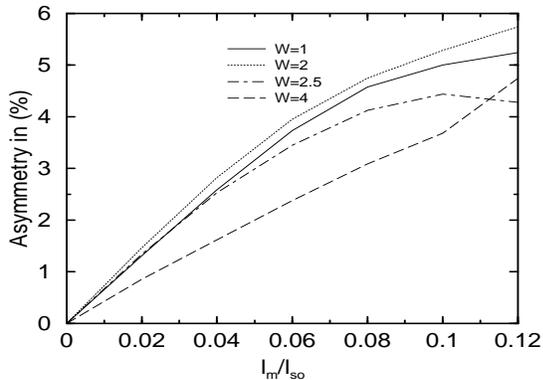,width=2.8in,height=1.5in,angle=0}}
\end{center}
\caption{ Spin asymmetry $A$, as a function of $l_{m}/l_{so}$
for different disorder strength.
The other parameters are
$\epsilon_f=3.0|t|$ and $\Delta=2.4|t|$ corresponding value of polarization is
50$\%$.  } 
\label{Fig. 5}
\end{figure}

In summary we have developed a new model to take into account spin-
orbit scattering within the single-band tight-binding model. Using
this model we have done numerical calculation of magnetic
scanning tunneling microscopy with a non-magnetic tip. The
order of magnitude of the spin-asymmetry is about $5\%$, which is
in good agreement with the qualitative estimate given in \cite{bruno}, and the
effect is maximum when the device operates in the quasi-ballistic regime.
The spin-asymmetry of the present effect is
smaller than the one obtained in the spin-valve tunneling 
structures. However, it has some advantages. In particular
since the tip is non-magnetic, it is insensitive to an external magnetic field
. This allows one to study the domain structure as function of applied field.
Furthermore, the problem of the magnetostatic interaction between the
tip and the magnetic sample is avoided, which incase of a magnetic
tip would give rise to undesirable magnetic forces between the
tip and the sample and are likely to influence the domain structure.
Another important advantages of this technique is that by measuring
separately the currents $I_2$ and $I_3$ of the two tip terminals,
and by combining them appropriately, one can separate the weak magnetic contrast from the dominant topographic contrast: the sum $I_{2}+I_{3}$ depends
only on the topography, whereas the magnetic information is contained in the
difference $I_{2}+I_{3}$. Besides all these advantages it has an intrinsic
limitation that the only in-plane components can be studied and
also since multiple scattering diminishes the spin-asymmetry
effect, it is necessary that the device operates in a quasi-ballistic regime.
However to construct such a tip would be experimentally challenging.

\end{multicols}
\end{document}